\documentclass[twocolumn,aps,pra,showpacs,superscriptaddress,longbibliography,10pt]{revtex4-1} %

\usepackage[colorlinks=true,linkcolor=blue,citecolor=blue]{hyperref}
\usepackage{amsfonts}
\usepackage{subfigure}
\usepackage{amsmath}
\usepackage{txfonts}
\usepackage{amssymb}
\usepackage{amsbsy}
\usepackage{epsfig}
\usepackage{graphicx}
\usepackage{epstopdf}
\usepackage{mathdots}
\usepackage{color}
\usepackage{cleveref}

\begin{document}
\title{Non-Hermitian Bulk-Boundary Correspondence in Periodically Driven System}
\author{Yang Cao}
\affiliation{Department of physics, Jiangsu University, Zhenjiang, 212013, China}

\author{Yang Li}
\affiliation{Department of physics, Jiangsu University, Zhenjiang, 212013, China}

\author{Xiaosen Yang} \altaffiliation{yangxs@ujs.edu.cn}
\affiliation{Department of physics, Jiangsu University, Zhenjiang, 212013, China}

\date{\today}

\begin{abstract}
Bulk-boundary correspondence, connecting the bulk topology and the edge states, is an essential principle of the topological phases. However, the bulk-boundary correspondence is broken down in general non-Hermitian systems. In this paper, we construct one-dimensional non-Hermitian Su-Schrieffer-Heeger model with periodic driving that exhibits non-Hermitian skin effect: all the eigenstates are localized at the boundary of the systems, whether the bulk states or the zero and the $\pi$ modes. To capture the topological properties, the non-Bloch winding numbers are defined by the non-Bloch periodized evolution operators based on the generalized Brillouin zone. Furthermore, the non-Hermitian bulk-boundary correspondence is established: the non-Bloch winding numbers ($W_{0,\pi}$) characterize the edge states with quasienergies $\epsilon=0, \pi$. In our non-Hermitian system, a novel phenomenon can emerge that the robust edge states can appear even when the Floquet bands are topological trivial with zero non-Bloch band invariant, which is defined in terms of the non-Bloch effective Hamiltonian. We also show that the relation between the non-Bloch winding numbers ($W_{0,\pi}$) and the non-Bloch band invariant ($\mathcal{W}$): $\mathcal{W}= W_{0}- W_{\pi}$.
\end{abstract}

\maketitle
Topological materials have robust topological edge states and can be characterized by the bulk topological invariants which is defined in terms of the Bloch Hamiltonian on Brillouin zone(BZ) \cite{HasanRevModPhys2010, qiRevModPhys2011,ChiuRevModPhys2016, BansilRevModPhys2016, kanePhysRevLett2005, MoorePhysRevB2007, fuliangPhysRevLett2007}. The topological surface states are connected with the bulk topological classification by the bulk-boundary correspondence\cite{kanePhysRevLett2005,MoorePhysRevB2007,fuliangPhysRevLett2007,Bernevig1757}, an essential principle of the topological phases.

Recently, non-Hermitian systems, described by the non-Hermitian Hamiltonian, have been attracting much attention in many fields of physics \cite{benderPhysRevLett1998, huicaoRevModPhys2015, Berry2004, YoungwoonPhysRevLett2010, Diehlnaturephys2011,AloisNature2012, WiersigPhysRevLett2014,HosseinNature2017,GanainyNaturePhysics2018, leePhysRevX2014, BaoganPhysRevA2014, malzardPhysRevLett2015, zhenbonature2015,PhysRevLettLee2016, PhysRevLettLee2016, zhongwangPhysRevLett2018a, zhongwangPhysRevLett2018b, XiaoLeiNaturePhysics2020, imura2020bloch, XuPhysRevLett2017, MenkePRB2017,ZengPRA2017, McDonaldPRX2018, PhysRevBYoshida2018, ShenPhysRevLett2018b,PhysRevACarlstrom2018,XiongJPC2018, GongPhysRevX2018,ShenPRL2018, ChenYuPhysRevB2018,KunstPhysRevLett2018, KunstPhysRevLett2018,ShenPhysRevLett2018b, EdvardssonPhysRevB2019,LieuPRB2019,YoshidaPhysRevB2019ER, YoshidaPhysRevB2019SPE, bergholtz2019exceptional, chang2019entanglement,midtgaard2019constraints,Hckendorf2019,OkumaPRL2019, LonghiPRL2019, YangPRB2019, KawabataNatureCommunications2019, HochenPhysRevResearch2020,kouPhysRevB2020,kouPhysRevB2020b, zhesenPhysRevLett2020, HelbigNaturePhysics2020, zhaihuiNP2020, YoshidaAPS2020,BorgniaPRL2020, zengPRB2020, SukhachovPRR2020, ZhenQianPhysRevLett2020, HaoranPhysRevLett2020,yang2020exceptional}. In particular, the conventional bulk-boundary correspondence is broken in general non-Hermitian systems, which exhibits the non-Hermitian skin effect
\cite{zhongwangPhysRevLett2018a, zhongwangPhysRevLett2018b,MartinezPhysRevB2018,LonghiPhysRevResearch2019, chenshuPRB2019, JiangbinhybridPRL2019, haijunPhysRevB2019, liuPhysRevLett2019, PhysRevLettwangzhong2019a, LuoPRL2019, lee2019unraveling, li2020critical,lee2020ultrafast, haga2020liouvillian, yi2020nonhermitian, OkumaPhysRevLett2020,YucePRA2020,zengPRR2020, li2020critical, liu2020helical,LinhuPhysRevLett2020}.
The wave functions of the non-Hermitian systems with open boundary condition(OBC) do not extend over the bulk but are localized at the boundaries, contrast to the Hermitian systems. To depict the topological properties of the non-Hermitian systems, the BZ should be extended into the generalized Brillouin zone(GBZ) in complex plane \cite{zhongwangPhysRevLett2018a, zhongwangPhysRevLett2018b, weiyiPhysRevB2019,YokomizoPhysRevLett2019,PhysRevLettwangzhong2019b,zhang2019correspondence,LonghiPhysRevLett2020, LeePhysRevB2019, KawabataPRX2019,xuearxiv2020,yangarxiv2019,XiaoLeiNaturePhysics2020}. The deviation between the GBZ and the BZ results in the non-Hermitian skin effect and the breakdown of the Bloch bulk-boundary correspondence\cite{zhongwangPhysRevLett2018a, zhongwangPhysRevLett2018b, YokomizoPhysRevLett2019, PhysRevLettwangzhong2019b, weiyiPhysRevB2019,zhang2019correspondence, XiaoLeiNaturePhysics2020,imura2020bloch, wuhongPhysRevB2020, LonghiPhysRevLett2020}. The universal topological invariants are defined by the non-Bloch Hamiltonian based on GBZ. The non-Bloch topological invariants capture the topological properties of the non-Hermitian systems like the robust topological edge modes\cite{zhongwangPhysRevLett2018a, zhongwangPhysRevLett2018b, YokomizoPhysRevLett2019}, the unidirectional edge motion\cite{zhongwangPhysRevLett2018b,KawabataPRX2019,HckendorPRL2019} and so on. Thus, the non-Hermitian bulk-boundary correspondence is well defined for general non-Hermitian systems and has been experimentally observed in photons\cite{XiaoLeiNaturePhysics2020} and metamaterials\cite{HelbigNaturePhysics2020, HofmannPRR2020,ghatak2019observation}.

In addition, periodic driving offers a new fruitful platform for creating topological phenomena with high tunability, even to create the fundamentally new topological states without static counterparts\cite{LimPhysRevLett2008, Soltannaturephy2011, Lindnernaturephy2011, RudnerPhysRevX2013, Rechtsmannature2013, FruchartPhysRevB2016, DrivenRevModPhys2017, YaoPRB2017, XiaosenScientific2018, ZhouPRB2018, HeNatureCommunications2019, LiPRB2019, ZhouPRB2020, ZhangPRB2020, he2020floquetengineering}. For instance, the robust topological edge modes can appear at the boundary of the systems even though all the band invariants are zero. Due to the periodicity in time, there exist two types of robust topological edge modes: zero modes and $\pi$ modes, which also has no static counterpart\cite{Lindnernaturephy2011,RudnerPhysRevX2013}.

In this paper, we interplay the non-Hermiticity and the periodic driving in one-dimensional Su-Schrieffer-Heeger(SSH) model by constructing the non-Bloch topological invariants in GBZ \cite{zhongwangPhysRevLett2018a,XiaoLeiNaturePhysics2020,wuhongPhysRevB2020,liarxiv2020}. First, the frequency-domain quasienergy spectra are given both in the OBC and the periodic boundary condition(PBC). There is a sharp difference between the spectra in the two conditions. We find that the periodic driving induces two types of topological edge modes with quasienergies $\epsilon=0$ and $\epsilon=\pi$ in OBC. The robust edge modes cannot be predicted by the winding number defined on BZ. To capture the robust topological edge modes, we construct the non-Bloch winding numbers defined in terms of the non-Bloch periodized evolution operators based on GBZ. The non-Bloch winding numbers exactly predict the zero modes and the $\pi$ modes. Then, the non-Hermitian bulk-boundary correspondence is established for the non-Hermitian system with periodic driving. Secondly, we give the two equivalent definitions of non-Bloch band invariant defined in terms of the frequency-domain and the effective Hamiltonian on GBZ. We find that the robust edge modes can appear when the Floquet bands are topological trivial with zero non-Bloch band invariant. Lastly, we show that the deviation between the GBZ and BZ induces the non-Hermitian skin effect: all the eigenstates are localized at the boundaries of the system, whether the bulk states or the topological zero and $\pi$ modes. We also give the phase diagram of the system by the non-Bloch band theory and the spectra of Floquet Hamiltonian in OBC.

\begin{figure*}
\centering
\includegraphics[width=0.9\textwidth]{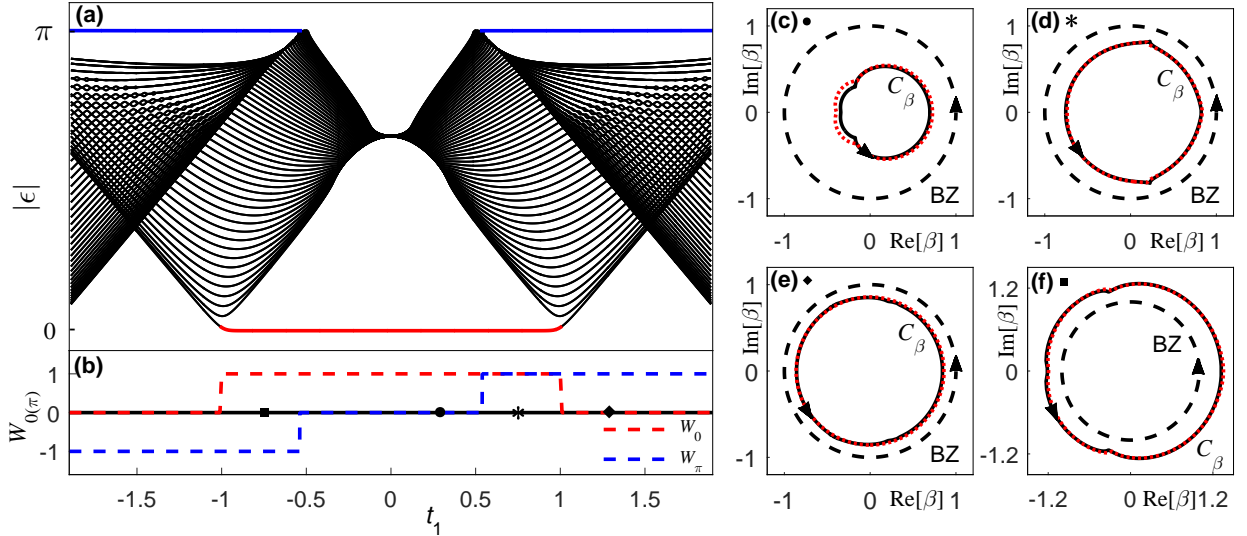}
\caption{\textbf{Non-Hermitian Bulk-Boundary Correspondence for Edge States with quasienergies $\epsilon =0,\pi$:} (a) Quasienergy spectrum in open boundary condition and (b) non-Bloch winding numbers with the parameters: $t_{2}=1$, $\gamma=0.2$, $\lambda=0.5$ and $\omega=3$. (c)-(f) The generalized Brillouin zone of the four marked topological different phases in (b) determined by non-Bloch Floquet Hamiltonian (solid) and non-Bloch effective Hamiltonian (dotted). The values of the parameters are (c) $t_{1} = 0.3$, (d) $t_{1} = 0.75$, (e) $t_{1} = 1.3$ and (f) $t_{1} = -0.75$. The Brillouin zone is a unit circle (dashed). The non-Bloch winding numbers ($W_{0,\pi}$), defined on the generalized Brillouin zone, characterize the zero and $\pi$ modes in open boundary condition.} \label{nhbbc}
\end{figure*}

\textit{Non-Hermitian Bloch Hamiltonian.--} We consider a non-Hermitian SSH model\cite{zhongwangPhysRevLett2018a} with periodic driving. The Bloch Hamiltonian can be written as following:
\begin{eqnarray}
H(k,t) = [d_{x} + \lambda \cos(\omega t)]\sigma_{x} + [d_{y} +i \gamma]\sigma_{y},
\label{Blochhamiltonian}
\end{eqnarray}
where $d_{x} = t_{1} + t_{2} \cos(k)$, $d_{y} = t_{2} \sin(k)$ and $\sigma_{x,y}$ are the Pauli matrices. $\gamma$ and $\lambda$ are the strengths of non-Hermitian and periodic driving respectively. The Hamiltonian has a chiral symmetry\cite{ChiuRevModPhys2016} $\sigma_{z} H(k,t) \sigma_{z}= - H(k, -t)$ and periodicity $H(k, t) = H(k, t + T)$ with periodic time $T=2 \pi/\omega$. The chiral symmetry ensures the quasienergies appear in $(E, -E)$ pairs.

\textit{Non-Hermitian Bulk-Boundary Correspondence.--} For simplicity, we numerical calculate the quasienergy spectra of the non-Hermitian Hamiltonian by the frequency-domain formulation\cite{Lindnernaturephy2011,RudnerPhysRevX2013}. Let us start from the time-dependent Schr$\ddot{\mathrm{o}}$dinger equation for the Floquet state in band $n$:
\begin{eqnarray}
i\partial_{t} \left|\psi_{n,R}(k,t)\right> = H(k,t) \left|\psi_{n,R}(k,t)\right>,
\end{eqnarray}
and take the Floquet theorem and the Fourier transformation to the frequency domain,
$\left|\psi_{n,R}(k,t)\right> = \mathrm{exp}[- i \varepsilon_{n}(k) t ] \sum_{m} \mathrm{exp}(i m \omega t ) \left|\psi_{n,R}^{m}(k)\right>$, where $\varepsilon_{n} $ is the quasienergy and $\left|\psi_{n,R}^{m}(k)\right>$ is right wave vector. In frequency domain framework, the Schr$\ddot{\mathrm{o}}$dinger equation becomes
\begin{eqnarray}
\sum_{m'} \mathcal{H}_{m,m'}(k) \left|\psi_{n,R}^{m'}(k)\right> = \varepsilon_{n}(k)\left|\psi_{n,R}^{m}(k)\right>, \label{floqh2}
\end{eqnarray}
with the Floquet Hamiltonian $\mathcal{H}_{m,m'}(k) = m \omega \delta_{m,m'} \mathbf{I} + H_{m-m'}(k)$ and $H_{m}(k) = \frac{1}{T} \int_{0}^{T}dt H(k,t) \mathrm{exp}(-i m\omega t)$. To be more explicit, the Floquet Hamiltonian reads
\begin{eqnarray}
\mathcal{H}=
\left(
  \begin{array}{ccccc}
    ... &        &      &      &   \\
        & H_{0}+ \omega & H_{1} & 0 &   \\
        &  H_{-1} & H_{0} & H_{1}  &  \\
        &  0 & H_{-1}  & H_{0} - \omega &  \\
        &   &   &    &   ... \\
  \end{array}
\right),
\end{eqnarray}
here, $H_{0}=d_{x}\sigma_{x} + [d_{y} +i \gamma]\sigma_{y}$ and $H_{\pm 1} = \frac{\lambda}{2} \sigma_{x}$. The eigenvalues of the Floquet Hamiltonian are the quasienergies of the non-Hermitian system. The quasienergies are paired with $(E,-E)$ for the chiral symmetry $\mathcal{C}^{-1} \mathcal{H} \mathcal{C} = -\mathcal{H}$\cite{YaoPRB2017}.

To show the quasienergy of the Floquet Hamiltonian $\mathcal{H}(k)$, we need to make a truncation on the Floquet Hamiltonian for its infinite rank. Fig.\ref{nhbbc}(a) shows the dimensionless quasienergy $\epsilon = \varepsilon T$ of Floquet Hamiltonian as functions of $t_{1}$ in OBC with $t_{2}=1$, $\gamma=0.2$, $\lambda=0.5$ and $\omega=3$. In the presence of periodic driving, unconventional gap can emerge around quasienergy $\epsilon=\pi$, which enables the stability of robust nontrivial $\pi$ modes. Thus, two topological different types of the nontrivial edge modes can exist at the two separated gaps with quasienergies $\epsilon=0,\pi$. This phenomena seeming the same as the cases of Hermitian systems, while they are utterly different in topological characterizing and non-Hermitian skin effect. In the non-Hermitian systems, the two types of robust edge modes can be characterized by the topological invariants not rely on the Bloch band theory like the Hermitian cases, but on the non-Bloch band theory in generally. Fig.\ref{nhbbc}(b) shows the non-Bloch winding numbers $W_{0,\pi}$ as functions of $t_{1}$, which are defined in terms of non-Bloch periodized evolution operators based on GBZ $C_{\beta}$ and can exactly characterize the numbers of the two robust topological different edge modes.

\textit{Generalized Brillouin Zone and Non-Bloch Winding Numbers.--} The precise description of the non-Hermitian systems relies on the non-Bloch band theory based on the GBZ. To determine the GBZ, we rewrite the Bloch Floquet Hamiltonian $\mathcal{H}(k)$ in terms of $\beta=e^{ik}$ as non-Bloch Floquet Hamiltonian $\mathcal{H}(\beta)$. Then, the eigenvalue equation $\det[\mathcal{H}(\beta) - E]=0$ is an algebraic equation for $\beta$ with even degree. Due to $2\pi$ modules of the  quasienergies $\epsilon$, we constrain the quasienergy in one period such as $\epsilon \in [-\pi, \pi]$. The solutions are numbered as $\beta_{i}(i=1,2,...,2N)$ and satisfy $|\beta_{1}| \leq |\beta_{2}| \leq ... \leq |\beta_{2N}|$. Thus, the GBZ $C_{\beta}$ can be determined by the trajectory of $\beta_{N}$ and $\beta_{N+1}$ with $|\beta_{N}|=|\beta_{N+1}|$\cite{zhongwangPhysRevLett2018a,YokomizoPhysRevLett2019,XiaoLeiNaturePhysics2020} as shown in Fig.\ref{nhbbc}(c)-(f) for four marked points in Fig.\ref{nhbbc}(b) with $t_{1}=0.3$, $0.75$, $1.3$ and $-0.75$. GBZ, always deviates from BZ in general, is smaller than BZ for Fig.\ref{nhbbc}(c)-(e) and larger than BZ for Fig.\ref{nhbbc}(f).

The non-Bloch time-evolution operator $U(\beta,t)$ can be got by rewriting the Bloch Hamiltonian $H(k,t)$ and is defined as
\begin{eqnarray}
U(\beta,t)= \mathcal{T} \mathrm{exp}\left[-i \int_{0}^{t} dt' H(\beta,t')\right],
\end{eqnarray}
where $\mathcal{T}$ is the time-ordering operator. The non-Bloch time-evolution operator satisfies the differential equation $i\partial_{t} U(\beta,t) = H(\beta,t) U(\beta,t)$. For one period, the non-Bloch Floquet operator $U(\beta,T)$ can be expanded as $U(\beta,T) =  \sum _{n} \lambda_{n}(\beta) | \psi_{n,R}(\beta) \rangle \langle \psi_{n,L}(\beta) |$. When there is a gap in the spectrum of the non-Bloch Floquet operator around $e^{-i \epsilon}$, we can define an unambiguous non-Bloch effective Hamiltonian as
\begin{eqnarray}
H_{\mathrm{eff}}^{\epsilon} (\beta) = \frac{i}{T} \mathrm{ln}_{- \epsilon} U(\beta,T).\label{enbh}
\end{eqnarray}
The subscript $\epsilon$ has been introduced as the branch cut which plays an essential ingredient in the construction of topological invariants for non-Bloch Floquet systems.  We take $\ln_{\epsilon} e^{i \phi} = i  \phi$ for $\epsilon - 2\pi < \phi < \epsilon$. Then, the effective Hamiltonian breaks the chiral symmetry with $\sigma_{z} H_{\mathrm{eff}}^{\epsilon} (\beta) \sigma_{z} = - H_{\mathrm{eff}}^{-\epsilon} (\beta) + \omega$ and can be rewritten as
$H_{\mathrm{eff}}^{\epsilon} (\beta) =  \frac{i}{T} \sum _{n} \mathrm{ln}_{- \epsilon} \left( \lambda_{n}(\beta) \right) | \psi_{n,R}(\beta) \rangle \langle \psi_{n,L}(\beta) |$.
The effective Hamiltonian only captures the stroboscopic evolution at the quasienergy of $\epsilon$ and lost the information of time-evolution within each period. So, the non-Bloch topological invariants are defined by the non-Bloch periodized evolution operator, which can be written as following:
\begin{eqnarray}
U_{\epsilon}(\beta,t)= U(\beta,t) \mathrm{exp}\left[i  H_{\mathrm{eff}}^{\epsilon} (\beta) t \right],
\end{eqnarray}
with periodicity $U_{\epsilon}(\beta,t + T) = U_{\epsilon}(\beta,t)$.  As for the chiral symmetry of the non-Bloch Hamiltonian, non-Bloch periodized evolution operator has the chiral symmetry only at half-period with quasienergy $\epsilon=0,\pi$:
\begin{eqnarray}
\sigma_{z} U_{0}(\beta,T/2)\sigma_{z} &=& - U_{0}(\beta,T/2),\nonumber\\
\sigma_{z} U_{\pi}(\beta,T/2)\sigma_{z} &=& + U_{\pi}(\beta,T/2).
\end{eqnarray}
Therefore, $U_{0}(\beta,T/2)$ and $U_{\pi}(\beta,T/2)$ are off-diagonal and diagonal respectively, which can be written as following:
\begin{eqnarray}
U_{0}(\beta,T/2)&=&
\left(
  \begin{array}{cc}
       0 & U_{0}^{+}(\beta) \\
      U_{0}^{-}(\beta)  &  0 \\
  \end{array}
\right),\\
U_{\pi}(\beta,T/2)&=&
\left(
  \begin{array}{cc}
      U_{\pi}^{+}(\beta)  &  0 \\
       0 &  U_{\pi}^{-}(\beta)  \\
  \end{array}
\right).
\end{eqnarray}

Then, the non-Bloch winding numbers of Yao-Wang formula can be written as following\cite{zhongwangPhysRevLett2018a,XiaoLeiNaturePhysics2020}:
\begin{eqnarray}
W_{\epsilon=0,\pi} = \frac{i}{2 \pi} \int_{C_{\beta}} \mathrm{Tr}\left[\left(U_{\epsilon}^{+}(\beta)\right)^{-1} d U_{\epsilon}^{+}(\beta)\right],
\label{nbwn}
\end{eqnarray}
which characterize the numbers of the Floquet zero and $\pi$ modes at the quasienergies of $\epsilon=0,\pi$ as shown in Fig.\ref{nhbbc}(b). Therefore, we construct the non-Hermitian bulk-boundary correspondence that the robust topological nontrivial zero and $\pi$ modes are characterized by the non-Bloch winding numbers ($W_{\epsilon=0,\pi}$).

\begin{figure}
\includegraphics[width=0.45\textwidth]{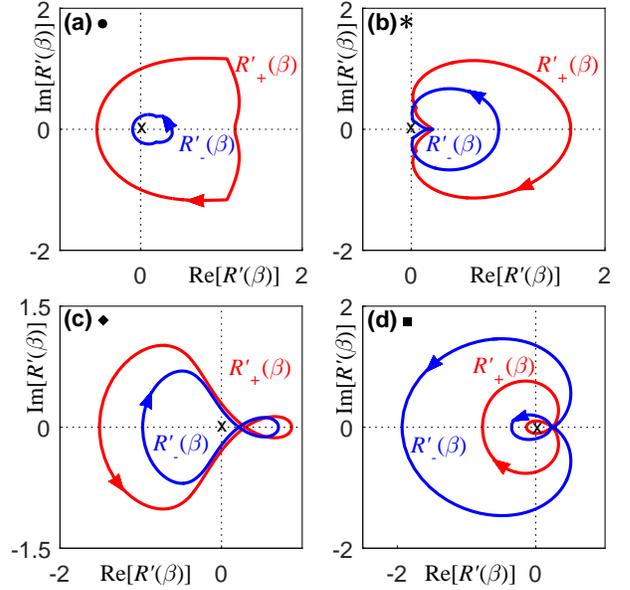}
\caption{(a)-(d) show the loops of $R'_{+} (\beta)$ (red) and $R'_{-} ( \beta)$ (blue) on the complex plane along the generalized Brillouin zone $C_{\beta}$ of Fig.\ref{nhbbc}(c)-(f), respectively. } \label{rloop}
\end{figure}

\textit{Non-Bloch Band Invariant.--} With the help of the Floquet operator $U(\beta,T)$, we can define the non-Bloch band invariant for the Floquet bands \cite{RudnerPhysRevX2013,YaoPRB2017,XiaoLeiNaturePhysics2020,wuhongPhysRevB2020,liarxiv2020}. We also can give the non-Bloch band invariant easily and intuitively. For single resonance case, the non-Bloch effective Hamiltonian can be derived from the Floquet Hamiltonian:
\begin{eqnarray}
H_{\mathrm{eff}} (\beta) = R'_{+}(\beta) \sigma_{+} + R'_{-}(\beta) \sigma_{-},
\end{eqnarray}
with
$R'_{\pm}(\beta) = \left[1- \frac{\omega}{2 d_{\beta}} \mp \frac{\lambda[R_{+}(\beta)-R_{-}(\beta)]}{4 d_{\beta}^{2}} \right] R_{\pm}(\beta)$,
$R_{\pm}(\beta) = t_{1} \pm \gamma + t_{2} \beta^{\mp}$ and $d_{\beta} = \sqrt{R_{+}(\beta) R_{-}(\beta)}$.
The GBZ $C_{\beta}$ can be derived by the eigenvalue equation $E^{2} = R_{+}^{'}(\beta) R_{-}^{'}(\beta)$ \cite{zhongwangPhysRevLett2018a,YokomizoPhysRevLett2019}, which is consistent with the one derived from the above non-Bloch Floquet Hamiltonian as shown in Fig.\ref{nhbbc}(c)-(f). The GBZs are determined by non-Bloch Floquet Hamiltonian (solid) and non-Bloch effective Hamiltonian (dotted) for the four topological different phases. The two curves coincide with each other.

\begin{figure}
\includegraphics[width=0.48\textwidth]{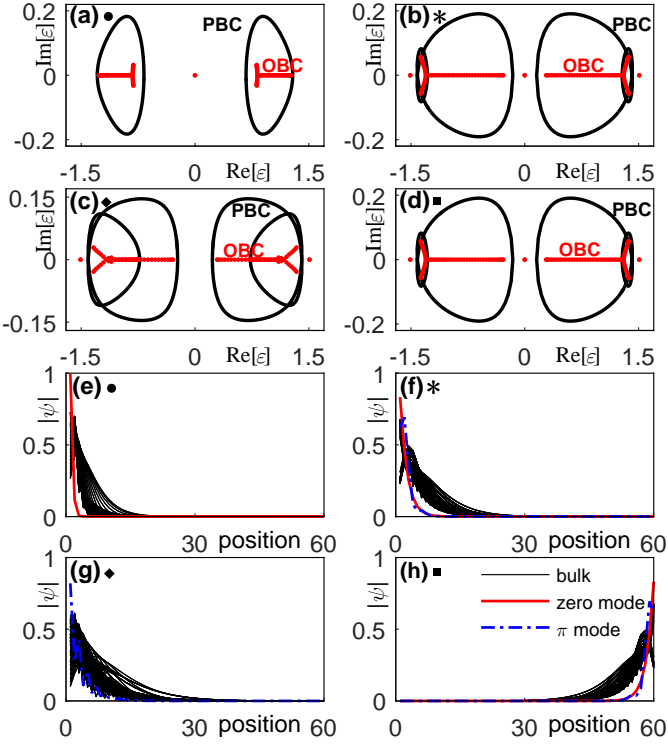}
\caption{\textbf{Non-Hermitian Skin Effect of Floquet System.} (a)-(d) The quasienergy spectra under periodic boundary condition (black) and open boundary condition (red) of the four topological different phases in Fig.\ref{nhbbc}(b). The values of the parameters are $\omega=3$ (a) $t_{1} = 0.3$, (b) $t_{1} = 0.75$, (c)$t_{1} = 1.3$ and (d) $t_{1} = -0.75$. (e)-(h) The bulk eigenstates (black) and the zero (red) and $\pi$ (blue) modes of the Floquet Hamiltonian in open boundary condition for (a)-(d), respectively.} \label{se}
\end{figure}

Due to the chiral symmetry of the effective Hamiltonian $\sigma_{z} H_{\mathrm{eff}}(\beta) \sigma_{z} = - H_{\mathrm{eff}} (\beta) $, the generalized 'Q matrix' $Q(\beta)$ is given by
\begin{eqnarray}
Q(\beta) = \frac{1}{ \sqrt{R'_{+}(\beta) R'_{-}(\beta)}} \left(
  \begin{array}{cc}
       0 &   R'_{+}(\beta)    \\
      R'_{-}(\beta)  &  0  \\
  \end{array}
\right).
\label{Qmat}
\end{eqnarray}
Thus, we obtain $q = R'_{+}(\beta)/\sqrt{R'_{+}(\beta) R'_{-}(\beta)}$ and have the Yao-Wang formula of the non-Bloch band invariant\cite{zhongwangPhysRevLett2018a,YokomizoPhysRevLett2019} for Floquet bands as:
\begin{eqnarray}
\mathcal{W} &=& \frac{i}{2 \pi} \int_{C_{\beta}} d q ~ q^{-1}(\beta) \nonumber\\
 &=&-\frac{1}{4\pi} \left[ \arg R'_{+}(\beta) - \arg R'_{-}(\beta) \right]_{C_{\beta}}.
\label{nbcn}
\end{eqnarray}

The non-Bloch band invariant is determined by the changes of the phases of $R'_{\pm}(\beta)$ when $\beta$ goes along GBZ $C_{\beta}$ in the counterclockwise way.
Fig.\ref{rloop}(a)-(d) show the loops of $R'_{+} ( \beta )$ (red) and $R'_{-} ( \beta )$ (blue) on the $R'$ plane along the GBZ of Fig.\ref{nhbbc} (c)-(f) in the counterclockwise way. Thus, the non-Bloch band invariants are $1, 0, -1$ and $2$ for Fig.\ref{rloop}(a)-(d), respectively. Fig.\ref{rloop}(b) shows that the original point is not encircled by $R'_{+} ( \beta )$ and $R'_{-} ( \beta )$, which indicates that the Floquet band is topological trivial ($\mathcal{W}=0$). While, the phase is topological nontrivial with nonzero non-Bloch winding numbers ($W_{0,\pi}=1$) and has topological robust zero and $\pi$ modes. Thus, the non-Bloch band invariant only characterizes the topological properties of the Floquet bands and cannot characterizes the robust topological edge modes. This is due to the Floquet operator also lost the information of time-evolution within each period. Fig.\ref{rloop}(d) shows that both of the $R'_{+} ( \beta )$ and $R'_{-} ( \beta )$ wind the origin point twice but in the opposite direction. The non-Bloch band invariant is $2$ with nontrivial non-Bloch winding numbers $W_{0}=1$ and $W_{\pi}=-1$. Certainly, the relation between the non-Bloch winding numbers and the non-Bloch band invariant is $W_{0} - W_{\pi} = \mathcal{W}$ (refer to Supplemental Materials). The non-Bloch winding numbers are more fundamental than the non-Bloch band invariant of the Floquet bands.

\textit{Non-Hermitian Skin Effect.--} We have shown that the GBZs deviate from the BZ in Fig.\ref{nhbbc}. Undoubtedly, the deviation induces not only the above breakdown of Bloch bulk-boundary correspondence but the non-Hermitian skin effect, which is another significantly different from Hermitian systems. Fig.\ref{se}(a)-(d) show the quasienergy spectra of the Floquet Hamiltonian in OBC(red) and PBC(black) for the four topological different phases in Fig.\ref{nhbbc}(b). The quasienergy spectra of the Floquet Hamiltonian have essential difference between the OBC and PBC, which is also induced by the deviation between the GBZ and BZ. Certainly, Floquet Hamiltonian $\mathcal{H}(\beta)$ can undoubtedly give the quasienergy spectra with OBC and PBC in Fig.\ref{se}(a)-(d) along the GBZ $C_{\beta}$ and the BZ in Fig.\ref{nhbbc}(c)-(f), respectively. This essential difference indicates the emergence of the non-Hermitian skin effect as shown in Fig.\ref{se}(e)-(h). Fig.\ref{se}(e)-(h) show the eigenstates of the Floquet Hamiltonian in OBC with lattice size $L=60$.  All the eigenstates of the non-Hermitian system are localized at the ends, which is called non-Hermitian skin effect. When $|\beta|$ is smaller than 1, the correspond eigenstate is localized at the left end of the system, Otherwise, the eigenstate is localized at the right end of the system\cite{PhysRevLettwangzhong2019b}. Accordingly, the eigenstates of the system are localized at the left ends for (e)-(g) and right ends for $(h)$, whether the eigenstates of bulk bands or the two topological different edge modes. This coincides with the non-Hermitian systems without periodic driving\cite{zhongwangPhysRevLett2018a}.

\begin{figure}
\includegraphics[width=0.45\textwidth]{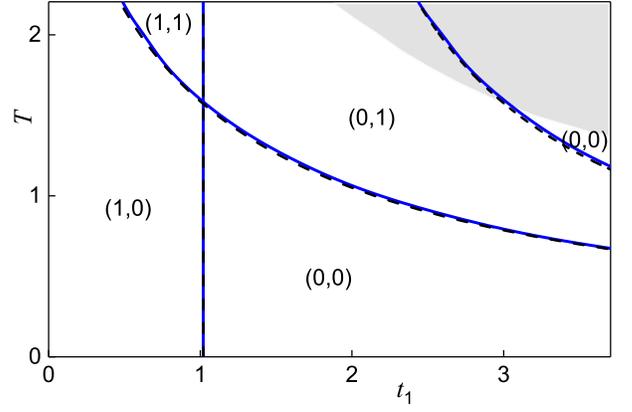}
\caption{\textbf{The Phase Diagram on the $T-t_{1}$ plane:} The phase diagram is determined by the non-Bloch band theory (dashed) and the quasienergy of Floquet Hamiltonian (solid) with $t_{2}=1$, $\gamma=0.2$ and $\lambda=0.5$. The shaded region represents that the non-Bloch topological invariants are not well-defined for the gapless quasienergy. Four topological different phases with non-Bloch winding numbers: $(W_{0}, W_{\pi}) = (1,0), (0,0),(0,1)$ and $(1,1)$. For $t_{1}<0$, only $W_{\pi}$ changes the sign.}  \label{pd}
\end{figure}
\textit{The Phase diagram.--} Based on the quasienergies of Floquet Hamiltonian in OBC and the non-Bloch band theory, we can draw the phase diagram on the $T-t_{1}$ plane with $t_{2}=1$, $\gamma=0.2$ and $\lambda=0.5$ as shown in Fig.\ref{pd}. For weak driving case, the phase boundaries are approximated as $t_{1} = \pm \sqrt{t_{2}^{2} + \gamma^{2}}$ and $ \omega = 2 \left|t_{2} \pm \sqrt{\left|t_{1}^{2}- \gamma^{2}\right|} \right|$, which can be determined by the changing of the non-Bloch band invariant. In the phase diagram, there are four topological different phases: (i) only having zero modes $(1,0)$, (ii) having any edge modes $(0,0)$, (iii) only having $\pi$ modes $(0,1)$ and (iv) having both zero and $\pi$ modes $(1,1)$. The Floquet bands are not isolated in the shaded region, where the non-Bloch topological invariants are not well defined.

{\it Conclusion:--}
We investigated the topological properties of the non-Hermitian Su-Schrieffer-Heeger model with periodic driving which exhibits non-Hermitian skin effect: all the eigenstates are localized at the edges. We found that there are two types of robust edge modes with quasienergies $\epsilon=0,\pi$. The topology of the edge modes cannot be characterized by the winding number defined on the Brillouin zone. Therefore, we established the non-Hermitian bulk-boundary correspondence: the robust edge modes can be characterized by the non-Bloch winding numbers defined by the non-Bloch Floquet operators on generalized Brillouin zone. The non-Hermitian bulk-boundary correspondence connects the bulk topology and the robust topological edge modes. In our system, we found the robust edge modes can appear when the Floquet bands are topological trivial with zero non-Bloch band invariant.

{\it Acknowledgment:--}
We thank  Zhong Wang for fruitful discussion. This work is supported by NSFC under Grants No.11504143.

\bibliography{reference}

\clearpage

{\bf Supplemental Materials}

\vspace{7mm}

{\it Symmetry:---}
The chiral symmetry of the non-Bloch Hamiltonian:
\begin{eqnarray}
\sigma_{z} H(\beta,t) \sigma_{z} = -H(\beta, -t).
\end{eqnarray}

The non-Bloch evolution operator:
\begin{eqnarray}
&& \sigma_{z} U(\beta,t) \sigma_{z}\nonumber\\
&=& \sigma_{z} \mathcal{T} \left[ e^{-i \int_{0}^{t} dt' H(\beta,t')} \right]\sigma_{z}\nonumber\\
&=&\sigma_{z} \left[ e^{ - i \Delta t H(\beta, t)} \cdots  e^{ - i \Delta t H(\beta, \Delta t)} \right] \sigma_{z}\nonumber\\
&=& \sigma_{z} [1 - i \Delta t H(\beta, t) ]\sigma_{z} \cdots   \sigma_{z}^{-1} [1 - i \Delta t H(\beta, \Delta t) ]\sigma_{z}\nonumber\\
&=& [1 + i \Delta t H(\beta, - t) ]  \cdots  [1 + i \Delta t H(\beta, -  \Delta t) ]\nonumber\\
&=& \left\{ [1 - i \Delta t H(\beta, -  \Delta t) ] \cdots  [1 - i \Delta t H(\beta, - t) ] \right\}^{-1}\nonumber\\
&=& U ^{-1}(\beta, t) = U (\beta, - t)
\end{eqnarray}

The effective non-Bloch Hamiltonian:
\begin{eqnarray}
 && \sigma_{z}H_{\mathrm{eff}}^{\epsilon} (\beta) \sigma_{z}\nonumber\\
&=& \frac{i}{T} \ln_{-\epsilon} \left[ \sigma_{z} U (\beta,  T) \sigma_{z} \right]\nonumber\\
&=& \frac{i}{T} \ln_{-\epsilon} \left[ U ^{-1} (\beta,  T)  \right]\nonumber\\
&=& \frac{i}{T} \sum_{n} \ln_{-\epsilon} \left( \lambda_{n}^{-1}\right) | \psi_{n,R}(\beta) \rangle \langle \psi_{n,L}(\beta) |  \nonumber\\
&=& \frac{i}{T} \sum_{n}  \left[ - \ln_{-\epsilon} (\lambda_{n}) - 2\pi i \right] | \psi_{n,R}(\beta) \rangle \langle \psi_{n,L}(\beta) |  \nonumber\\
&=& - H_{\mathrm{eff}}^{- \epsilon} (\beta) + \omega
\end{eqnarray}

The non-Bloch periodized evolution operator:
\begin{eqnarray}
&&\sigma_{z} U_{\epsilon} (\beta,t) \sigma_{z}\nonumber\\
&=& \sigma_{z}  U (\beta,t) \sigma_{z} \sigma_{z}  \mathrm{exp}\left[ i H_{\mathrm{eff}}^{\epsilon} (\beta) t \right] \sigma_{z}\nonumber\\
&=&  U (\beta, -t)  \mathrm{exp}\left[ i  \sigma_{z} H_{\mathrm{eff}}^{\epsilon} (\beta) \sigma_{z} t \right] \nonumber\\
&=& U (\beta, -t)  \mathrm{exp}\left[ - i   H_{\mathrm{eff}}^{-\epsilon} (\beta)  t + i \omega t \right] \nonumber\\
&=& U_{-\epsilon} (\beta, -t)  \mathrm{exp}\left[ i \omega t \right].
\end{eqnarray}

For $\epsilon = 0, \pi$, we have
\begin{eqnarray}
\sigma_{z} U_{0} \left(\beta, \frac{T}{2}\right) \sigma_{z} &=&  - U_{0} \left(\beta, \frac{T}{2}\right),\\
\sigma_{z} U_{\pi} \left(\beta, \frac{T}{2}\right) \sigma_{z} &=& + U_{\pi} \left(\beta, \frac{T}{2}\right).
\end{eqnarray}

Then, $U_{0,\pi}$ takes the following forms:
\begin{eqnarray}
U_{0}(\beta,T/2)&=&
\left(
  \begin{array}{cc}
       0 & U_{0}^{+}(\beta) \\
      U_{0}^{-}(\beta)  &  0 \\
  \end{array}
\right),\\
U_{\pi}(\beta,T/2)&=&
\left(
  \begin{array}{cc}
      U_{\pi}^{+}(\beta)  &  0 \\
       0 &  U_{\pi}^{-}(\beta)  \\
  \end{array}
\right).
\end{eqnarray}
Then,
\begin{eqnarray}
U_\pi ^{-1}\left(\frac{T}{2}\right) U_0\left(\frac{T}{2}\right) =
\left(
  \begin{array}{cc}
       0  &  \left[U_{\pi}^{+}\right]^{-1}U_{0}^{+} \\
       \left[U_{\pi}^{-}\right]^{-1}U_{\pi}^{-} &  0  \\
  \end{array}
\right).
\end{eqnarray}

{\it Relationship between the non-Bloch winding numbers and non-Bloch Band invariant:---} We prove the relation between the non-Bloch winding numbers and the the non-Bloch band invariant $\mathcal{W}(P_{0,\pi}) = W\left[U_0 \left(\beta, \frac{T}{2}\right) \right] - W\left[U_\pi \left(\beta, \frac{T}{2}\right) \right]$ in the main text. The difference between the two non-Bloch effective Hamiltonian is:
\begin{eqnarray}
 &&H_{\mathrm{eff}}^{\pi} (\beta) -  H_{\mathrm{eff}}^{0} (\beta)\nonumber\\
&=& \frac{i}{T} \mathrm{ln}_{- \pi} U(\beta,T) - \frac{i}{T} \mathrm{ln}_{0} U(\beta,T)\nonumber\\
&=& \frac{i}{T} \sum_{n} \left[\ln_{-\pi} \left( \lambda_{n}\right) - \ln_{0} \left( \lambda_{n}\right) \right]| \psi_{n,R}(\beta) \rangle \langle \psi_{n,L}(\beta) | \nonumber\\
&=& \frac{i}{T} \left( \sum_{-\pi<\epsilon_{n}<0} +  \sum_{0<\epsilon_{n}<\pi} \right) \left[\ln_{-\pi} \left( e ^{-i\epsilon_{n}}\right) - \ln_{0} \left( e ^{-i\epsilon_{n}} \right) \right]| \psi_{n,R}(\beta) \rangle \langle \psi_{n,L}(\beta) |\nonumber\\
&=& \frac{i}{T} \sum_{0<\epsilon_{n}<\pi} \left[-i\epsilon_{n}- 2 \pi i + i\epsilon_{n} \right]| \psi_{n,R}(\beta) \rangle \langle \psi_{n,L}(\beta) | \nonumber\\
&=& \frac{2\pi}{T} \sum_{0<\epsilon_{n}<\pi} | \psi_{n,R}(\beta) \rangle \langle \psi_{n,L}(\beta) | \nonumber\\
&=&\omega P_{0,\pi}(\beta).
\end{eqnarray}

The non-Bloch periodized evolution operator at half period:
\begin{eqnarray}
&&U_\pi ^{-1}\left(\beta, \frac{T}{2}\right) U_0\left(\beta, \frac{T}{2}\right)\nonumber\\
&=& \mathrm{exp}\left[ - i  H_{\mathrm{eff}}^{\pi} (\beta) \frac{T}{2} \right] U(\beta, -\frac{T}{2}) U(\beta, \frac{T}{2}) \mathrm{exp}\left[ i  H_{\mathrm{eff}}^{0} (\beta) \frac{T}{2} \right] \nonumber\\
&=& \exp \left(- i \pi P_{0,\pi}\right) \nonumber\\
&=& 1 - 2 P_{0,\pi}\nonumber\\
&=& Q(\beta).
\end{eqnarray}

For the winding number, there is additive property:
\begin{eqnarray}
W\left[U_\pi ^{-1}\left(\beta, \frac{T}{2}\right) U_0\left(\beta, \frac{T}{2}\right)\right] = W\left[U_0 \left(\beta, \frac{T}{2}\right) \right] - W\left[U_\pi \left(\beta, \frac{T}{2}\right) \right]. \nonumber
\end{eqnarray}

\begin{figure}
\includegraphics[width=0.45\textwidth]{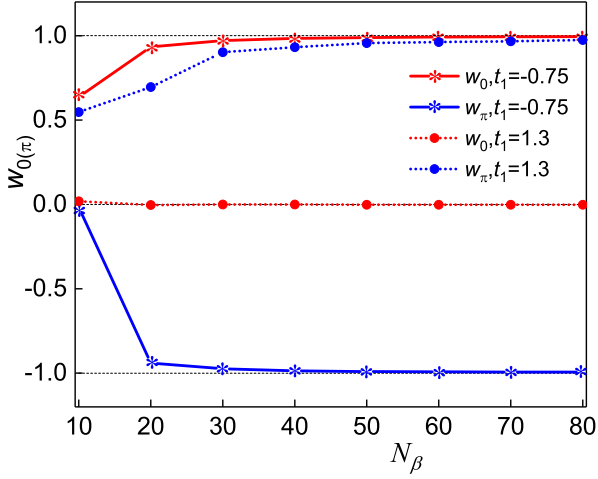}
\caption{Non-Bloch winding numbers $W_{0}$ and $W_{\pi}$ as functions of the GBZ size ($N_{\beta}$) for $t_{1}=-0.75$(solid) and $t_{1}=1.3$(dotted), with $t_{2}=1$, $\gamma=0.2$, $\lambda=0.5$, $\omega=3$. Increasing the size, the non-Bloch winding numbers quickly approach into an integer.}  \label{wn}
\end{figure}

Therefore, the relationship between the non-Bloch winding numbers and the non-Bloch band invariant can be written as following:
\begin{eqnarray}
W\left[U_0 \left(\beta, \frac{T}{2}\right) \right] - W\left[U_\pi \left(\beta, \frac{T}{2}\right) \right] = \mathcal{W}.
\end{eqnarray}

{\it Topological Phase Boundaries Based on the Effective Hamiltonian:-}
The non-Bloch effective Hamiltonian is defined by the non-Bloch Floquet operator:
\begin{eqnarray}
H_{\mathrm{eff}} (\beta) &=& \frac{i}{T} \mathrm{ln} U(\beta,T),\nonumber\\
&=& H_{0}(\beta) + \sum_{n\neq0} \frac{[H_{0}, H_{n}]}{n \omega} + \sum_{n > 0}\frac{[H_{n}, H_{-n}]}{n\omega}.
\end{eqnarray}
Here, $H_{0}(\beta) = R_{+}(\beta) \sigma_{+} + R_{-}(\beta) \sigma_{-}$ and $H_{\pm 1}(\beta)= \frac{\lambda}{2} \sigma_{x}$ with $R_{\pm}(\beta) = t_{1} \pm \gamma + t_{2} \beta^{\mp}$.

For high frequency case ($\omega > \Lambda$, $\Lambda$ is the bandwidth of $H_{0}$), the effective Hamiltonian is not effected by the periodic driving and can be written as $H_{\mathrm{eff}}(\beta)= R'_{+}(\beta) \sigma_{+} + R'_{-}(\beta) \sigma_{-}$ with $R'_{\pm}(\beta) = R_{\pm}(\beta)$.  The generalized Brilouin zone is a circle $C_{\beta} = \sqrt{\left|\frac{t_{1} -\gamma}{t_{1} + \gamma}\right|} e^{i \theta}$ with $\theta = [0, 2\pi)$.

For low frequency case ($\omega < \Lambda$), the drive creates a single resonance between the two bulk bands of $H_{0}$. Therefore, the non-Bloch effective Hamiltonian $H_{\mathrm{eff}}(\beta)$ is defined by the rotated Floquet operator $U(\beta,T) = \mathcal{T} \mathrm{exp}\left[-i \int_{0}^{T} dt~H_{\mathrm{rot}}(\beta,t)\right]$. The rotated Hamiltonian $H_{\mathrm{rot}}(\beta,t)$ can be deduced by applying a rotation $\hat{O}(\beta,t)= \exp \left[- i \hat{\mathbf{d}}_{\beta} \cdot \mathbf{\sigma} \omega t /2 \right]$ to the non-Bloch Hamiltonian $H(\beta,t)$ with $\hat{\bf{d}}_{\beta}= {\bf d}_{\beta}/ d_{\beta}$ with $d_{\beta}=\sqrt{d_{x}^{2}+d_{y}^{2}}$. Then, the non-Bloch effective Hamiltonian  $H_{\mathrm{eff}}(\beta)$ can be rewritten as following:
\begin{eqnarray}
H_{\mathrm{eff}} (\beta) = R'_{+}(\beta) \sigma_{+} + R'_{-}(\beta) \sigma_{-},
\end{eqnarray}
with
\begin{eqnarray}
R'_{\pm}(\beta) = \left[1- \frac{\omega}{2 d_{\beta}} \mp \frac{\lambda[R_{+}(\beta)-R_{-}(\beta)]}{4 d_{\beta}^{2}} \right] R_{\pm}(\beta).
\end{eqnarray}

For weak driving, the formula of $R'$ can be rewritten as
\begin{eqnarray}
R'_{\pm}(\beta) \simeq \left[1- \frac{\omega}{2 d_{\beta}} \right] R_{\pm}(\beta).
\end{eqnarray}

Then, the generalized Brilouin zone can be determined and written as
\begin{eqnarray}
C_{\beta} = \sqrt{\left|\frac{t_{1} -\gamma}{t_{1} + \gamma}\right|} e^{i \theta},
\end{eqnarray}
with $\theta = [0, 2\pi)$.

The gap closing equation can be derived directly:
$R_{+}(\beta) R_{-}(\beta) =0$ and $\omega = 2 d_{\beta}$.

Thus, the topological phase boundaries can be written as following:
\begin{eqnarray}
t_{1} &=& \pm \sqrt{t_{2}^{2} + \gamma^{2}}\\
\omega &=& 2 \left|t_{2} \pm \sqrt{\left|t_{1}^{2}- \gamma^{2}\right|} \right|.
\end{eqnarray}

\end{document}